\documentclass[letterpaper, 10 pt, conference]{ieeeconf}  

\IEEEoverridecommandlockouts                              
\usepackage{cite}
\usepackage{amsmath}
\usepackage{amsfonts}
\usepackage{amssymb}
\usepackage{subfig}
\usepackage{epsfig}
\usepackage{graphicx}
\usepackage{multicol}
\title{\LARGE \bf
A Mean Value Theorem Approach to Robust Control Design for Uncertain Nonlinear Systems
}

\author{Obaid Ur Rehman  \qquad Ian R. Petersen \qquad  Bar\i\c{s} Fidan
\thanks{O.Rehman  is a student of Electrical Engineering,
        University of New South Wales at Australian defence force Academy, 2610 ACT, Australia
        {\tt\small s.obaid.rehman@gmail.com}}%
\thanks{I.R.Petersen  is a professor in School of Engineering and Information Technology,
        University of New South Wales at Australian defence force Academy, 2610 ACT, Australia
        {\tt\small i.r.petersen@gmail.com}}
        \thanks{B. Fidan is an assistant professor at Mechanical and Mechatronics Engineering Department, University of Waterloo, N2L3G1 ON, Canada
{\tt\small fidan@uwaterloo.ca}}%
}
\begin{document}
\newtheorem{definition}{Definition}
\newtheorem{remark}{Remark}
\newtheorem{theorem}{Theorem}
\newcommand{\finishproof}{\hfill $\blacksquare$}
\newcommand{\startproof}{\vspace{0.2cm} \noindent {\bf \em Proof: }}
\maketitle
\thispagestyle{empty}
\pagestyle{empty}

\begin{abstract}
This paper presents a  scheme to design a tracking controller for a class of uncertain nonlinear systems using a robust feedback linearization approach.  The scheme is composed of two steps. In the first step, a linearized uncertainty model for the corresponding uncertain nonlinear system is developed using a robust feedback linearization approach.  In this step, the standard feedback linearization approach is used to linearize the nominal nonlinear dynamics of the uncertain nonlinear system. The remaining nonlinear uncertainties are then linearized at an arbitrary point using the mean value theorem. This approach gives a multi-input multi-output (MIMO) linear uncertain system model with a structured uncertainty representation. In the second step, a minimax linear quadratic regulation (LQR) controller is designed for MIMO linearized uncertain system model. In order to demonstrate the  effectiveness of the proposed method, it is applied to a velocity and altitude tracking control problem for an air-breathing hypersonic flight vehicle.
\end{abstract}
\section{Introduction}
In this paper, a robust tracking control scheme is designed for a class of uncertain nonlinear systems.  The design is composed of two steps. In the first step a linearized uncertainty model for the uncertain nonlinear system is developed using a robust feedback linearization approach.
The feedback linearization approach has many applications in the process control and aerospace industries. Using this method, a large class of nonlinear systems can be made to exhibit linear input-output behavior using a nonlinear state feedback control law. Once the input-output map is linearized, any linear controller design method can be used to design a desired controller. One of the limitations of the standard feedback linearization method is that the model of the system must be exactly known. In the presence of uncertainty in the system, exact feedback linearization is not possible and uncertain nonlinear terms remain in the system.

In order to resolve the issue of  uncertainty after canceling the nominal nonlinear terms using the feedback linearization method, several approaches have been considered in the literature \cite{FBRC,NBSS,FBLetter,FBchemical,FBRC,FBbound,FB_robust_Friedovich,FBobserver1}. Most of these approaches use adaptive or related design procedures to estimate the uncertainty in the system. In these methods,  mismatched uncertainties are decomposed into matched and mismatched parts. These methods typically require the mismatched parts not to exceed some maximum allowable  bound \cite{FBoutput}. The existing results that are based on mismatched uncertainties either do not guarantee stability or require some stringent conditions on the nonlinearities and uncertainties to guarantee the stability of the system \cite{FBchemical,FBRC}.

In this paper, we approach the uncertainty issue in a different way and represent the uncertain nonlinear system in an uncertain linearized form. We use a nominal feedback linearization method to cancel the nominal nonlinear terms, and use a generalized mean value theorem to linearize the nonlinear uncertain terms. In our previous work \cite{Rehman_ASJC,CDC02}, the uncertain nonlinear terms are linearized using a Taylor expansion at a steady state operating point by considering a structured representation of the uncertainties. This linearization approach approximates the actual nonlinear uncertainty by considering only the first order terms and neglecting all of the higher order terms.  In \cite{CDC02}, the uncertain nonlinear terms are linearized using a Taylor expansion but an unstructured representation of uncertainty is considered. In both of these methods, the linearized uncertainty model was obtained by ignoring higher order terms.  In \cite{Rehman_ASCC01}, we introduced the linearization of nonlinear terms using using a generalized mean value theorem \cite{Lin_algebra,Mcleod_Mean_value} approach. This method exactly linearizes the uncertain nonlinear terms at an arbitrary point and therefore, no higher order terms exist. In \cite{Rehman_ASCC01}, the upper bound on the uncertainties is obtained by using unstructured uncertainty representations. The bound obtained using an unstructured uncertainty representation may be conservative which may degrade the performance of the closed loop system. In order to reduce conservatism  and to obtain an uncertain linearized model with a structured uncertainty representation, a different approach for obtaining an upper bound is presented here. In contrast to \cite{Rehman_ASCC01}, here we propose a minimax linear quadratic regulation (LQR) \cite{IP} controller which combines with a standard feedback linearization law and gives a stable closed loop system in the presence of uncertainty. Here, we assume that the uncertainty satisfies a certain integral quadratic constraint (IQC).

The paper is organized as follows. Section \ref{sec:PSlqr} presents a description of the considered class of uncertain nonlinear systems. Our approach to robust feedback linearization is given in Section \ref{sec:FBlqr}. Derivation of the linearized uncertainty model and tracking controller for an air-breathing hypersonic flight vehicle (AHFV) along with simulation results are presented in Section \ref{sec:VM}. The paper is concluded in Section \ref{sec:concl} with some final remarks on the proposed scheme.

\section{System Definition}\label{sec:PSlqr}
Consider a multi-input multi-output (MIMO) uncertain nonlinear system with the same number of inputs and outputs as follows:
\begin{align}
\label{eqsystemlqr}
\dot{x}(t)&=f(x,\hat{p})+\sum\limits_{k=1}^m{g_k(x,\hat{p})u_k(t)}+\epsilon \bar{g}(\hat{p},x,u),\nonumber\\
y_i(t)&=\nu_i(x),\quad i=1,2,\cdots,m
\end{align}
where $x(t)\in \mathbb{R}^n$, $u(t)=[u_1.....u_m]^T\in\mathbb{R}^m$, $y(t)=[y_1....y_m]^T\in\mathbb{R}^m$ and $\epsilon\neq0$.
Furthermore, the system has norm bounded uncertain parameters lumped in the vector $\hat{p}\in \mathbb{R}^q$. Also, $f(x,\hat{p})$,  $g_i(x,\hat{p})$, and  $\nu_i(x,\hat{p})$ for $i=1,\cdots,m$ are assumed to be infinitely differentiable (or differentiable to a sufficiently large degree) functions of their arguments. The term $\bar{g}(\hat{p},x,u)$ in (\ref{eqsystemlqr}) is a nonlinear function which represents the couplings in the system.  The full state vector $x$ is assumed to be available for measurement.

\section{Feedback Linearization}\label{sec:FBlqr}
In this section, we first simplify the model (\ref{eqsystemlqr}) so that the term involving $\epsilon$ vanishes. Here we assume that $\vert\epsilon\vert$ is sufficiently small and hence indicates weak coupling. In general, $\epsilon$ depends on the physical parameters of the system and may be available through measurement or known in advance. Instead of neglecting this coupling in the control design, we model $\epsilon \bar{g}(\hat{p},x,u)$ as an uncertainty function $\tilde{g}(\hat{p},\bar{p},x)$ with certain bound, where, $\bar{p}$ denotes a new uncertainty parameter whose magnitude is bounded. The parameter $\bar{p}$ appears due to the removal of coupling terms which depend on the input $u$. Now we can write (\ref{eqsystemlqr}) as follows:
\begin{equation}
\label{eqSSsystem}
\begin{split}
\dot{x}(t)&=\bar{f}(x,p)+\sum\limits_{k=1}^m{g_k(x,\hat{p})u_k(t)}\\
y_i(t)&=\nu_i(x),\quad i=1,2,\cdots,m\\
\end{split}
\end{equation}
where $\bar{f}(x,p)=f(x,\hat{p})+\tilde{g}(\hat{p},\bar{p},x)$ is an infinitely differentiable function and $p=[\hat{p} \quad \bar{p} ]^T$. Also, note that in equation (\ref{eqSSsystem}), which includes the new uncertain parameter $\bar{p}$, we can write the system in terms of a new uncertainty  vector $p=p_0+\Delta p \in \mathbb{R}^{\bar{q}}$, where  $\bar{q}=q+1$. Here, $p_0$ is the vector of the nominal values of the parameter vector $p$ and $\Delta p$ is the vector of uncertainties in the corresponding parameters as follows:
\begin{align*}
&p_{0}=\left[\begin{array}{ccccc}
p_{10} & p_{20} & \cdots &  p_{({\bar{q}}-1)0} & p_{{\bar{q}}0}
\end{array}\right],\\
&\Delta p=\left[\begin{array}{ccccc}
\Delta p_1 & \Delta p_2 & \cdots &  \Delta p_{({\bar{q}}-1)} & \Delta p_{\bar{q}}
\end{array}\right].
\end{align*}
We assume that a bound on $|\Delta p_s|$ is known for each $s\in \{1,\cdots,\bar{q}\}$. We also assume that the functions in the system (\ref{eqSSsystem}) are differentiable. The standard feedback linearization method can be used on the nominal model (without uncertainties) by differentiating each individual element $y_i$ of the output vector $y$ a sufficient number of times until a term containing the control element $u$ appears explicitly. The number of differentiations needed is equal to the relative degree $r_i$ of the system with respect to each output for $i=1,2,~\cdots~m$. Note that a nonlinear system of the form (\ref{eqSSsystem}) with $m$ output channels has a vector relative degree $r=[r_1~r_2~\cdots~r_m]$ \cite{IS}. We assume that the nonlinear system (\ref{eqSSsystem}) has full relative degree; i.e. $\sum\limits_{i=1}^m{r_i}=n$, where $n$ is the order of the system.

It is shown in \cite{CDC01} that in the presence of uncertainties exact cancellation of the nonlinearities is not possible because only an upper bound on the uncertainties is known: Indeed, we obtain
\begin{equation}
\label{eqSsystem}
\begin{split}
\dot{x}(t)&=\underbrace{\bar{f}_0(x,p_0)+\sum\limits_{k=1}^m{g_{k0}(x,\hat{p_0})u_k(t)}}_\text{Nominal part}\\
&+\underbrace{\Delta \bar{f}(x,p)+\sum\limits_{k=1}^m{\Delta g_k(x,p)u_k(t)}}_\text{Uncertain part}\\
y_i(t)&=\nu_i(x),\quad i=1,2,\cdots,m
\end{split}
\end{equation}
where, $\Delta \bar{f}$, and $\Delta g_k$ are the uncertain parts of their respective functions. After taking the Lie derivative of the regulated outputs a sufficient number of times, the system (\ref{eqSsystem})  can be written as follows:
\begin{align}
\label{eqdiffoutputlqr}
\left[\begin{array}{c}
y_1^{r_1}\\
\vdots \\
y_m^{r_m}
\end{array}\right]
&=f_*(x)+g_*(x)u \nonumber\\
&+\left[\begin{array}{c}
L_{\Delta\bar{f}}^{r_1}(\nu_1)+\sum\limits_{k=1}^m L^{r_1-1}_{ \Delta g_k}[L_{ \Delta \bar{f}}( \nu_1)]u_k\\
\vdots\\
L_{\Delta\bar{f}}^{r_m}(\nu_m)+\sum\limits_{k=1}^m L^{r_m-1}_{\Delta g_k}[L_{\Delta \bar{f}}(\nu_m)]u_k
\end{array}\right],
\end{align}
where,
\begin{footnotesize}
\begin{align*}
f_*(x)&=[L_{\bar{f}_0}^{r_1}(\nu_1) \cdots L_{\bar{f}_0}^{r_m}(\nu_m)]^T,\\
g_*(x)&=\left[
\begin{array}{cc}
L_{g_{10}}L_{\bar{f}_0}^{r_1-1}(\nu_1)& L_{g_{20}}L_{\bar{f}_0}^{r_1-1}(\nu_1)\dots L_{g_{m0}}L_{\bar{f}_0}^{r_1-1}(\nu_1)\\
L_{g_{10}}L_{\bar{f}_0}^{r_2-1}(\nu_2) & L_{g_{20}}L_{\bar{f}_0}^{r_2-1}(\nu_2)\dots L_{g_{m0}}L_{\bar{f}_0}^{r_2-1}(\nu_2)\\
\vdots &\vdots\quad\quad\quad\quad\quad\quad\vdots \\
L_{g_{10}}L_{\bar{f}_0}^{r_m-1}(\nu_m)&L_{g_{20}}L_{\bar{f}_0}^{r_m-1}(\nu_m)\dots L_{g_{m0}}L_{\bar{f}_0}^{r_m-1}(\nu_m)
\end{array}\right],
\end{align*}
\end{footnotesize}
and the Lie derivative of the functions $\nu_i$ with respect to the vector fields $\bar{f}$ and $g_k$ are given by
\begin{footnotesize}
\begin{align*}
L_{\bar{f}} \nu_i=&\frac{\partial \nu_i (x)}{\partial x} \bar{f} ,~ L_{\bar{f}}^j\nu_i=L_{\bar{f}} (L_{\bar{f}}^{j-1}\nu_i (x))
,~ L_{g_k }(\nu_i)=\frac{\partial \nu_i (x)}{\partial x}g_k.
\end{align*}
\end{footnotesize}
Note that in equation (\ref{eqdiffoutputlqr}), we have deliberately lumped the uncertainties at the end of a chain of integrators. This is because the uncertainties in $y^{1}_i,\cdots,y^{r_i-1}_i$ will be included in the diffeomorphism, which will be defined in the sequel. This definition of the diffeomorphism is in contrast to \cite{CDC01}, where the uncertainties in $y^{1}_i,\cdots,y^{r_i-1}_i$ are assumed to be zero; i.e. they satisfy a generalized matching condition\cite{NBSS}.

The feedback control law
\begin{equation}
\label{equ}
u=-g_*(x)^{-1}f_*(x)+g_*(x)^{-1}v,
\end{equation}
partially linearizes the input-output map (\ref{eqdiffoutputlqr}) in the presence of uncertainties as follows:

\begin{align}
\label{eqsdiffoutputm}
&y^{r_i}_*=\underbrace{\left[
\begin{array}{c}
v_1\\
\vdots\\
v_m
\end{array}\right]}_\text{Nominal part}+
\underbrace{\left[
\begin{array}{c}
\Delta W_1^{r_1}(x,u,p)\\
\vdots\\
\Delta W_m^{r_m}(x,u,p)
\end{array}\right]}_\text{Uncertainty part},
\end{align}
where
$\Delta W_i^{r_i}(x,u,p)=L_{\Delta \bar{f}}^{r_i}( \nu_i)+\sum\limits_{k=1}^m L^{r_i-1}_{\Delta g_k}[L_{\bar{f}}( \nu_i)]u_k$,  $y_*=[y_1^{r_1} .... y_m^{r_m}]^T$, and $v=[v_1 .... v_m]^T$ is the new control input vector. Furthermore, we define an uncertainty vector \begin{small}$\Delta W_i$\end{small} which represents the uncertainty in each derivative of the $i^{th}$ regulated output as
\begin{equation}
\label{eqWi}
\begin{split}
\Delta W_i(x,u,p)
&=\left[\begin{array}{c}
L_{\Delta \bar{f}}(\nu_i)\\
L_{\Delta \bar{f}}^{2}(\nu_i)\\
\vdots\\
L_{\Delta \bar{f}}^{r_i}( \nu_i)+\sum\limits_{k=1}^m L^{r_i-1}_{\Delta g_k}[L_{\bar{f}}( \nu_i)]u_k
\end{array}\right]\\&=\left[\begin{array}{c}
\Delta W_i^{1}(x,u,p)\\
\Delta W_i^{2}(x,u,p)\\
\vdots\\
\Delta W_i^{r_i}(x,u,p)
\end{array}\right],
\end{split}
\end{equation}
and write $y_i$ for $i=1,2,\cdots,m$ as given below.
\begin{small}
\begin{equation}
\label{eqyi}
\left[\begin{array}{c}
y_i^1\\
y_i^2\\
\vdots\\
y_i^{r_i}
\end{array}\right]=\left[\begin{array}{c}
0\\
0\\
\vdots\\
v_i
\end{array}\right]+\left[\begin{array}{c}
\Delta W_i^{1}(x,u,p)\\
\Delta W_i^{2}(x,u,p)\\
\vdots\\
\Delta W_i^{r_i}(x,u,p)
\end{array}\right].
\end{equation}
\end{small}
Let us define a nominal diffeomorphism similar to the one defined in \cite{Rehman_ASCC01} for each partially linearized system in (\ref{eqyi}) for $i=1,\cdots, m$ as given below:
\begin{equation}
\label{eqdiffm}
\chi_i=T_i(x,p_0)=\left[\begin{array}{ccccc}
\int y_i-y_{ic} & y_i-y_{ic} & y^{1}_i & .. & y^{r_i-1}_i\end{array}\right]^{T}.
\end{equation}
Using the diffeomorphism (\ref{eqdiffm}) and system (\ref{eqyi}), we obtain the following:
\begin{equation}
\label{eqfullBvsky}
\dot{\chi}=A{\chi}+B v+\Delta \bar{W}(\chi,v,p),\quad\quad
\end{equation}
where $\chi(t)=[\chi_1^T(t),\cdots,\chi_m^T(t)]^T\in \mathbb{R}^{\bar{n}}$ ($\bar{n}=n+m$), $v(t)=[v_1~v_2 \cdots v_m]^T\in\mathbb{R}^m$ is the new control input vector, $ \Delta \bar{W}(\chi,v,p)=[\Delta \bar{W}_1^T(.), \Delta \bar{W}_2^T(.),~ \cdots ~,  \Delta \bar{W}_m^T(.)]^T$ is a transformed version of $\Delta W (x,u,p)$ using (\ref{eqdiffm}) and $\Delta\bar{W}_i(.)=[w_i^{(1)}(.), w_i^{(2)}(.),~ \cdots,~ w_i^{(r_i)}(.)]^T$ for $i=1,2,~\cdots,~m$. Also,
\[
A=\left[
\begin{array}{ccc}
A_1&  \dots  & 0\\
\vdots   & \ddots & \vdots \\
0 & \dots & A_m
\end{array}\right];
~~B=\left[
\begin{array}{cccc}
\bar{B}_1&  \dots  & 0\\
\vdots   & \ddots & \vdots \\
0 &  \dots & \bar{B}_m
\end{array}\right],
\]
where
\[
A_i=\left[\begin{array}{cccccc}
0 & 1 & 0 & \cdots & 0 & 0\\
0 & 0 & 1 & \cdots & 0 & 0\\
\vdots & &\vdots & &\vdots &\\
0 & 0 & 0 & 0 & 0 & 0\\
\end{array}\right],\quad
\bar{B}_i=\left[\begin{array}{c}
0\\
\vdots
\\
0\\
1\\
\end{array}\right].
\]

In our previous work, these uncertainty terms are linearized at a steady state operating point and all the higher order terms in states, control and parameters are ignored in order to obtain a fully linearized form for (\ref{eqfullBvsky}). In this paper, we adopt a different approach to the linearization of the uncertain nonlinear terms in (\ref{eqfullBvsky}). Here, we perform the linearization of $\Delta \bar{W}(\chi,v,p)$ using a generalized mean value theorem \cite{Lin_algebra, vector_valued_mean} such that no higher order terms exist.
\begin{theorem}\label{th_mean}
Let $\bar{w}_i^{(j)}:\mathbb{R}^{\bar{n}} \rightarrow \mathbb{R}$ be a differentiable mapping on $\mathbb{R}^{\bar{n}}$ with a Lipschitz continuous gradient $\nabla \bar{w}_i^{(j)}$. Then for given $\chi$ and $\chi(0)$ in $\mathbb{R}^{\bar{n}}$, there exists a vector $c_i=\chi+\bar{t}(\chi-\chi(0))$ with $\bar{t}\in [0,1]$, such that
\begin{equation}
\label{eqvect_mean}
\bar{w}_i^{(j)}(\chi)-\bar{w}_i^{(j)}(\chi(0)))=\nabla \bar{w}_i^{(j)}(c_i).(\chi-\chi(0)).
\end{equation}
\end{theorem}
\startproof
For proof, see \cite{vector_valued_mean}.
\finishproof


We can apply Theorem \ref{th_mean} to the nonlinear uncertain part of (\ref{eqfullBvsky}). Let us define a hyper rectangle
\begin{equation}
\mathfrak{B}=\{\left[
\begin{array}{c}
\chi\\
v
 \end{array}\right]:\begin{array}{c}
\underline{\chi_i}\leq\chi_i\leq\bar{\chi_i}\\
\underline{v_i}\leq v_i\leq\bar{v_i}
 \end{array}\},
\end{equation}
where $\underline{\chi_i}$, and $\underline{v_i}$  denote the lower bounds and $\bar{\chi_i}$, and $\bar{v_i}$ denote the upper bounds on the new states and inputs respectively. For this purpose, the gradient of $w_i^{(j)}(.)$ is found by differentiating it with respect to $\chi$ and $v$ at an arbitrary operating point $c_{ij}=(\tilde{\chi}, ~ \tilde{v}, ~ \tilde{p})$ for $i=1,2,\cdots,m$ and $j=1,2,\cdots, r_i$ where, $\left[\tilde{\chi}^T~~ \tilde{v}^T\right]^{T}\in\mathfrak{B}$, and $\tilde{p}\in\Theta$. We assume $w_i^{(j)}(\chi_0,v_0,p_0)=0 $, $\chi(0)=0$, and $v(0)=0$ and write $w_i^{(j)}(.)$ as follows:
\begin{equation}
w_i^{(j)}(\chi,v,p)  = {w'}_i^{(j)}(c_{ij}).[\chi^T \quad v^T]^T.
\end{equation}
 Then $\Delta \bar{W}(.)$ can be written as
\begin{equation}
\label{eqbarW}
\Delta\bar{W}(.)= \Phi \left[
\begin{array}{c}
\chi\\
v
 \end{array}\right],
\end{equation}
where
\[
\Phi=\left[
\begin{array}{c}
{w'}_{1}^{(1)}(c_{11}) \\
\vdots\\
 {w'}_{1}^{(r_1)}(c_{1r_1}) \\
\vdots\\
 {w'}_{m}^{(1)}(c_{m1}) \\
\vdots\\
 {w'}_{m}^{(r_m)}(c_{m r_m})
\end{array}\right].
\]
\subsection{Linearized model with structured uncertainty representation}\label{sec:structure_model}
The equation (\ref{eqfullBvsky}) can be written in a linearized form using (\ref{eqbarW}). Note that the matrix $\Phi$ is unknown. However, it is possible to write bounds on each term ${w'}_{i}^{(j)}(c_{ij})$ in $\Phi$ individually and represent them in a structured form. For this purpose, we define each individual bound as follows:
\begin{equation}
\label{eqrhos}
\begin{split}
\hat{\rho}_{1}&=\max_{c_{11}}(\Vert {w'}_{1}^{(1)}(c_{11}) \Vert),\\
\hat{\rho}_{2}&=\max_{c_{12}}(\Vert {w'}_{1}^{(2)}(c_{12}) \Vert), \\
&\vdots\\
\hat{\rho}_{r_1}&=\max_{c_{1r_1}}(\Vert {w'}_{1}^{(r_1)}(c_{1r_1})\Vert), \\
\hat{\rho}_{r_{1+1}}&=\max_{c_{21}}(\Vert {w'}_{2}^{(1)}(c_{21}) \Vert), \\
&\vdots\\
\hat{\rho}_{r_1+r_2+\cdots+r_m}&=\max_{c_{m r_m }}(\Vert {w'}_{m}^{(r_m)}(c_{m r_m }) \Vert).
\end{split}
\end{equation}
Using the definition in (\ref{eqrhos}) and (\ref{eqbarW}), the model (\ref{eqfullBvsky}) can be rewritten as
\begin{footnotesize}
\begin{equation}
\label{eqgforms}
\begin{split}
\dot {\chi}(t) &=(A+\tilde{C}_{1} \Delta_{1} \tilde{K}_{1}+\tilde{C}_{2} \Delta_{2} \tilde{K}_{2}+\cdots +\tilde{C}_{r_1} \Delta_{r_1} \tilde{K}_{r_1}+\cdots\\
&+\tilde{C}_{\bar{n}} \Delta_{\bar{n}} \tilde{K}_{\bar{n}})\chi(t)
+(B+\tilde{C}_{1} \Delta_{1} \tilde{G}_{1}+\tilde{C}_{2} \Delta_{2} \tilde{G}_{2}+\cdots \\
&+\tilde{C}_{r_1} \Delta_{r_1} \tilde{G}_{r_1}+\cdots
+\tilde{C}_{\bar{n}} \Delta_{\bar{n}} \tilde{G}_{\bar{n}}) v(t),
\end{split}
\end{equation}
\end{footnotesize}
where $\tilde{C}_{k}$ for $k=1,2,\cdots,\bar{n}$ is a $\bar{n} \times 1$  vector  whose $k$th entry is one and the other entries are zeros, $\tilde{K}_{k}$ for $k=1,2,\cdots,\bar{n}$ is a $1 \times \bar{n}$  vector whose $k$th entry is $\hat{\rho}_{k}$ and the other entries are zero,
$\tilde{G}_{1},\cdots,\tilde{G}_{r_{i-1}},\tilde{G}_{r_{i+1}},\cdots, \tilde{G}_{\bar{n}-1}=0$ for $i=1,2,\cdots,m$ and $\tilde{G}_{(.)}$ is a $1 \times m$  vector as defined below:
\begin{footnotesize}
\begin{equation}
\label{eqGs}
\begin{array}{c}
\tilde{G}_{r_1}=[\hat{\rho}_{r_1}~ \hat{\rho}_{r_1} \cdots \hat{\rho}_{r_1}],\\
\tilde{G}_{r_1+r_2}=[\hat{\rho}_{r_1+r_2}~ \hat{\rho}_{r_1+r_2} \cdots \hat{\rho}_{r_1+r_2}],\\
\vdots\\
\tilde{G}_{r_1+r_2+\cdots+r_m}=[\hat{\rho}_{r_1+r_2+\cdots+r_m}~ \hat{\rho}_{r_1+r_2+\cdots+r_m} \cdots \hat{\rho}_{r_1+r_2+\cdots+r_m}],
\end{array}
\end{equation}
\end{footnotesize}
and $\Vert \Delta_{k} \Vert < 1$.
Using the above definitions of variables, we will write the system in a general MIMO form as given below:
\begin{equation}
\label{eqggforms}
\begin{split}
\dot {\chi}(t) &=A\chi(t)+Bv(t)+\sum^{\bar{n}}_{k=1} \tilde{C}_k\zeta_k(t);\\
z_k(t)&=\tilde{K}_k{\chi(t)}+\tilde{G}_kv(t);\quad k=1,2,\cdots,\bar{n},\\
\end{split}
\end{equation}
where $\chi(t)\in \mathbb{R}^{\bar{n}}$ is the state; $\zeta_k (t)=\Delta_{k}[\tilde{K}_k\chi+\tilde{G}_1 v] \in \mathbb{R}$ is the uncertainty input, $z_k(t)\in \mathbb{R}$ is the uncertainty output, $v(t)\in\mathbb{R}^m$ is the new control input vector and $y(t)\in \mathbb{R}^m$ is the measured output vector.

\begin{theorem}
\label{th1m}
Consider the nonlinear uncertain system (\ref{eqsystemlqr}) with vector relative degree $\{r_1,\cdots,r_m\}$ at $x=x(0)$. Suppose also that $\bar{f}(x(0),p)=0$ and $\nu(x(0))=0$. There exist a feedback control law of the form (\ref{equ}) and coordinate transformation (\ref{eqdiffm}), defined locally around $x(0)$ transforming the nonlinear system into an equivalent linear controllable system (\ref{eqggforms}) with  uncertainty norm bound $\hat{\rho}$ for (\ref{eqggforms}) in a certain domain of attraction if
\begin{enumerate}
\item $L_{\Delta \bar{f}}L_{\bar{f}}^jh\neq0$ and $L_{\Delta g}L_{\bar{f}}^jh\neq0$ for all $0\leq i \leq r_i-1$,
\item $r_1+r_2+\cdots+r_m=n$,
\item the matrix $\beta =[g_*(x,p_0)]^{-1}$ exists,
\item The uncertainty satisfies $\vert \Delta \vert \leq 1 \quad \forall {\chi}\in \mathfrak{B}, {v}\in \mathfrak{B} \text{, and}~ p \in \Theta$.
\end{enumerate}
\end{theorem}
\startproof
The proof directly follows from the form of the feedback control law (\ref{equ}) which cancels all the nominal nonlinearities and the linearize the remaining uncertain nonlinear terms using the generalized mean value theorem at $\left[\tilde{\chi}~ \tilde{v}\right]^T\in\mathfrak{B}$, and $\tilde{p}\in \Theta$. Since the generalized mean value theorem allows us to write any nonlinear function as an equivalent linear function, which will be tangent to the nonlinear function at some given points, we can linearize the remaining uncertain nonlinear terms using the generalized mean value theorem. Finally, it is straight forward to write the entire uncertain nonlinear system (\ref{eqsystemlqr}) in the linear controllable form (\ref{eqggforms}) by finding the maximum norm bound  $\hat{\rho}$ in (\ref{eqrhos}) on the linearized uncertain terms in the region being considered, $\left[\tilde{\chi}~ \tilde{v}\right]^T\in\mathfrak{B}$, and ${p}\in \Theta$.
\finishproof

\section{AHFV Example}\label{sec:VM}
\subsection{Vehicle Model}
The nonlinear equations of motion of an AHFV used in this study are taken from the work of Bolender et al \cite{BD01} and the description of the coefficients are taken from Sigthorsson and Serrani \cite{LPV01}. The equations of motion are given below:
\begin{small}
\begin{equation*}
\label{eqn1}
\dot {V}=\frac{T \cos \alpha -D}{m}- g \sin\gamma,~~
\dot {\gamma }=\frac{L+T\sin \alpha }{mV}-\frac{g \cos \gamma }{V}
\end{equation*}
\begin{equation}
\label{eqn2}
\dot {h}=V\sin \gamma,\quad
\dot {\alpha }=Q-\dot {\gamma },~~
\dot {Q}=M_{yy} /I_{yy}
\end{equation}
\begin{equation*}
\label{eqn3}
\ddot{\eta_i}=-2\zeta_{m} w_{m,i} \dot{\eta_i}-w_{m,i}^2 \eta_i+N_i,\qquad i=1,2,3
\end{equation*}
\end{small}
See \cite{LPV01,BD01} for a full description of the variables in this model. The forces and moments in actual nonlinear model are intractable and do not give a closed representation of the relationship between control inputs and controlled outputs. In order to obtain tractable expressions for the purpose of control design, these forces and moments are replaced with curve-fitted approximations in \cite{LPV01} which leads to  a curve-fitted model (CFM). The CFM has been derived by assuming a flat earth and unit vehicle depth and retains the relevant dynamical features of the actual model and also offers the advantage of being analytically tractable \cite{LPV01}. The approximations of the forces and moments are given as follows in \cite{LPV01}:
\begin{small}
\begin{equation}
L\approx \bar{q} S C_L (\alpha ,\delta_e,\delta_c,\Delta \tau_1,\Delta \tau_2 ),
\end{equation}
\begin{equation}
D\approx \bar{q} S C_D (\alpha ,\delta_e,\delta_c,\Delta \tau_1,\Delta \tau_2 ),
\end{equation}
\begin{equation}
M_{yy} \approx z_T T+\bar{q} S \bar {c} C_M(\alpha ,\delta_e,\delta_c,\Delta \tau_1,\Delta \tau_2 ),
\end{equation}
\begin{equation}
T\approx \bar{q}  [\phi C_{T,\phi}(\alpha,\Delta \tau_1,M_{\infty})+C_T(\alpha,\Delta \tau_1,M_{\infty},A_d)],
\end{equation}
\begin{equation}
N_i\approx\bar{q} C_{N_i} [\alpha ,\delta_e,\delta_c,\Delta \tau_1,\Delta \tau_2],\quad i=1,2,3.
\end{equation}
\end{small}
The coefficients obtained from fitting the curves are given as follows. These coefficients are obtained by assuming states and inputs are bounded and only valid for the given range. Here, we remove the function arguments for the sake of brevity:
\begin{small}
\[
C_L=C_L^\alpha \alpha+C_L^{\delta_e} \delta_e+C_L^{\delta_c} \delta_c+C_L^{\Delta\tau_1}\Delta\tau_1+C_L^{\Delta\tau_2}\Delta\tau_2+C_L^0,\]
\[
C_M=C_M^\alpha \alpha+C_M^{\delta_e} \delta_e+C_M^{\delta_c} \delta_c+C_M^{\Delta\tau_1}\Delta\tau_1+C_M^{\Delta\tau_2}\Delta\tau_2+C_M^0,
\]
\begin{align*}
C_D =& C_D^{(\alpha+\Delta\tau_1)^2} (\alpha+\Delta\tau_1)^2+C_D^{(\alpha+\Delta\tau_1)} (\alpha+\Delta\tau_1)\\
&+C_D^{\delta_e^2} \delta_e^2+C_D^{\delta_e} \delta_e+C_D^{\delta_c^2} \delta_c^2+C_D^{\delta_c} \delta_c
+C_D^{\alpha\delta_e}\alpha\delta_e\\
&+C_D^{\alpha\delta_c}\alpha\delta_c+C_D^{\delta\tau_1} \delta\tau_1+C_D^0,
\end{align*}
\begin{align*}
C_{T,\phi}&=C_{T,\phi}^\alpha \alpha+C_{T,\phi}^{\alpha M_{\infty}^{-2}} \alpha M_{\infty}^{-2}+ C_{T,\phi}^{\alpha \Delta\tau_1} \alpha \Delta\tau_1\\
&+C_{T,\phi}^{M_{\infty}^{-2}} M_{\infty}^{-2}+C_{T,\phi}^{{\Delta\tau_1}^2} {\Delta\tau_1}^2+C_{T,\phi}^{{\Delta\tau_1}}{\Delta\tau_1}+C_{T,\phi}^0,
\end{align*}
\[
C_T=C_T^{A_d}A_d+C_T^\alpha \alpha+C_{T}^{M_{\infty}^{-2}} M_{\infty}^{-2}+C_{T}^{{\Delta\tau_1}}{\Delta\tau_1}+C_T^0,
\]
\[
C_{N_i}=C_{N_i}^\alpha \alpha+C_{N_i}^{\delta_e} \delta_e+C_{N_i}^{\delta_c} \delta_c+C_{N_i}^{\Delta\tau_1}\Delta\tau_1+C_{N_i}^{\Delta\tau_2}\Delta\tau_2+C_{N_i}^0.
\]
\end{small}
Here, $M_{\infty}$ is the free-stream Mach number, and $\bar{q}$ is the dynamic pressure, which are defined as follows:
\begin{small}
\begin{equation}
\bar{q}=\frac{\rho(h) V^2}{2},\quad
M_{\infty}=\frac{V}{M_0}.
\end{equation}
\end{small}
Also, $\rho(h)$ is the altitude dependent air-density and $M_0$ is the speed of sound at a given altitude and temperature.
The nonlinear equations of motion have five rigid body states; i.e., velocity $V$, altitude $h$, angle of attack $\alpha$, flight path angle $\gamma$, and pitch rate $Q$. The CFM also has $6$ vibrational modes and they are represented by generalized modal coordinates $\eta_i$. There are four inputs and they are the diffuser-area-ratio $A_d$, throttle setting or fuel equivalence ratio $\phi$, elevator deflection $\delta_e$, and canard deflection $\delta_c$. In this example, tracking of velocity and altitude will be considered.
\subsection{Feedback linearization of the AHFV nonlinear model}\label{sec:CLUM}
\subsubsection{Simplification of the CFM}\label{sec:SM}
The CFM contains input coupling terms in the lift and drag coefficients. We simplify the CFM in a robust way as presented in Section \ref{sec:FBlqr} so that the simplified model approximates the real model and the input term  vanishes in the low order derivatives during feedback linearization.
In the simplification process, we will first remove the flexible states as they have stable dynamics. A canard is introduced in the AHFV model by Bolender and Doman \cite{BD02} to cancel the elevator-lift coupling using an ideal interconnect gain $k_{ec}=-\frac{C_L^{\delta_e}}{C_L^{\delta_c}}$  which relates the canard deflection to elevator deflection ($\delta_c=k_{ec} \delta_e$). In practice, an ideal interconnect gain is hard to achieve and thus exact cancellation of the lift-elevator coupling is not possible. In the simplified model, we assume that the interconnect gain is uncertain with a bound on its magnitude and it  also satisfies an IQC \cite{IP}. The drag coefficient is also affected due to the presence of elevator and canard coupling terms in the corresponding expression. We also model this coupling as uncertainty.
The simplified expressions for lift, moment, drag, and  thrust coefficients now can be written as follows:
\begin{footnotesize}
\begin{equation}
\begin{split}
C_L&=C_L^\alpha \alpha+C_L^0+\Delta C_l,\\
C_M&=C_M^\alpha \alpha+[C_M^{\delta_e}-C_M^{\delta_c}(\frac{C_L^{\delta_e}}{C_L^{\delta_c}})] \delta_e+C_M^0,\\
C_D &= C_D^{(\alpha+\Delta\tau_1)^2} (\alpha)^2+C_D^{(\alpha+\Delta\tau_1)} (\alpha)+C_D^0+\Delta C_d,\\
C_{T,\phi}&=C_{T,\phi}^\alpha \alpha+C_{T,\phi}^{\alpha M_{\infty}^{-2}} \alpha M_{\infty}^{-2}\alpha+C_{T,\phi}^{M_{\infty}^{-2}} M_{\infty}^{-2}+C_{T,\phi}^0,\\
C_T&=C_T^{A_d}A_d+C_T^\alpha \alpha+C_{T}^{M_{\infty}^{-2}} M_{\infty}^{-2}+C_T^0,
\end{split}
\end{equation}
\end{footnotesize}
where $\Delta C_l$ is the uncertainty in the lift coefficient $C_L(\alpha)$ due to the uncertain interconnect gain and $\Delta C_d$ is the uncertainty in the drag coefficient $C_D(\alpha)$ due to the input coupling terms. Furthermore, in order to obtain full relative degree for the purpose of feedback linearization, we dynamically extend the system by introducing second order actuator dynamics into the fuel equivalence ratio input as follows:
\begin{small}
\begin{equation}
\label{eqfuel}
\ddot {\phi }=-2\zeta \omega _n \dot {\phi }-\omega _n^2 \phi +\omega
_n^2 \phi _c.
\end{equation}
\end{small}
After this extension we have two more states $\phi$ and $\dot{\phi}$, and thus the sum of the vector relative degree is equal to the order of the system $n$; i.e. $n=7$ and thus satisfying one of the conditions for feedback linearization\cite{IS}.
\begin{table}
\caption{States, Inputs and Selected Physical Parameters at the Trim Condition}
\label{tab01}
\centering
\begin{tabular}{l c}
\hline\hline
Vehicle Velocity $V$ & $7702.08$ ft/sec\\
Altitude $h$& 85000 ft \\
Fuel-to-air ratio $\phi$ & $0.4388$  \\
Pitch Rate $Q$ & $0$  \\
Angle of Attack $\alpha$ & $-0.0134$ rad \\
Flight Path Angle $\gamma$ & $0$\\
\hline \hline
Elevator Deflection $\delta_e$& $2.005$ deg\\
Canard Deflection $\delta_c$& $2.79$ deg \\
Diffuser Area ratio $A_d$ & $1$ \\
\hline \hline
Reference Area $S$ & $17$ sq-ft.ft$^{-1}$\\
Mean Aerodynamic Chord  $c$ & $17$ ft\\
Air Density $\rho$ & $6.6\times 10^{-5}$~slugs/ft$^3$\\
Mass with 50\% fuel level $m$ & $147.9$ slug. ft$^{-1}$ \\
Moment of Inertia $I_{yy}$ &  $5.0\times 10^{5}$~slugs/ft$^2$/(rad . ft)\\
\hline \hline
\end{tabular}
\end{table}
We use Theorem \ref{th1m} to linearize the AHFV dynamics. The outputs to be regulated are selected as velocity $V$ and altitude $h$ using two inputs, elevator deflection $\delta_e$, and fuel equivalence ratio $\phi_c$. Note that the canard deflection $\delta_c$ is a function of the elevator deflection $\delta_e$ and they are related via an interconnect gain. Also, we fix the diffuser area ratio $A_d$ to unity.
The new simplified model consists of seven rigid states $x=\left[\begin{array}{ccccccccc}
 V & h & \gamma & \alpha & \phi & \dot{\phi} & Q
\end{array}\right]^T$ and can be represented by a general form as follows:
\begin{small}
\begin{align}
\label{equsystemm}
\dot{x}(t)&=\bar{f}(x,p)+\sum\limits_{k=1}^2{g_k(x,p)u_k};\quad
y_i(t)=\nu_i(x),~i=1,2.
\end{align}
\end{small}
where the control vector $u$ and output vector $y$ are defined as
\begin{small}
\[
u=[u_1, u_2]^T=\left[ {\delta _e, \phi _c } \right]^T, y=[y_1, y_2]^T=\left[ {V,h}
\right]^T.
\]
\end{small}
The following set of uncertain parameters are considered for the development of a linearized uncertainty model:
\begin{small}
\begin{align}
p=[
&C_L^\alpha \quad C_M^{\delta_c} \quad C_{T,\phi}^{\alpha M_{\infty}^{-2}} \quad C_{T,\phi}^{M_{\infty}^{-2}} \Delta C_l \quad \Delta C_d \quad \Delta C_T \nonumber\\
&\Delta C_M \quad \Delta C_{T,\phi}]^T \in \mathbb{R}^{9}.
\end{align}
\end{small}
We assume that  $p \in \Theta$, where $\Theta=\{p\in \mathbb{R}^{9}~ |~ 0.9 p_0 \leq p_i \leq 1.1 p_0~~ \text{for}\quad i=1,\cdots,9\}$.
In order to get the linearized uncertain model for the uncertain nonlinear AHFV model, the output $V$ and the output $h$ are differentiated three times and four times respectively using the Lie derivative:
\begin{small}
\begin{equation}
\label{equv3dotm}
\left[
\begin{array}{c}
\stackrel{\dots}{V} \\
h^{4}
\end{array}\right]\underbrace{
=f_*(x,p_0)+
g_*(x,p_0)
\left[
\begin{array}{c}
\delta _e \\
\phi _c
\end{array}\right]}_\text{Nominal nonlinear part}+\underbrace{
\left[\begin{array}{c}
\Delta\stackrel{\dots}{V}(x,u,p) \\
\Delta {h}^4(x,u,p)\end{array}\right]}_\text{Uncertain nonlinear part},
\end{equation}
\end{small}
where
\begin{footnotesize}
\begin{equation*}
f_*(x,p_0)=\left[
\begin{array}{c}
L_f^3 V\\
L_f^4 h
\end{array}\right], \quad
g_*(x,p_0)=\left[\begin{array}{cc}
{L_{g_1}(L_f^2 V)} & {L_{g_2}(L_f^2 V)}\\
{L_{g_1}(L_f^3 h)} & {L_{g_2}(L_f^3 h)}
\end{array}\right],
\end{equation*}
\end{footnotesize}
and $\Delta\stackrel{\dots}{V}(x,u,p_0,\Delta p)$ and $\Delta {h}^4(x,u,p_0,\Delta p)$ are the uncertainties in $\stackrel{\dots}{V}$ and ${h}^4$ respectively.
The application of the control law (\ref{equ}) yields the following:
\begin{small}
\begin{equation}
\label{equfblin}
\left[\begin{array}{c}
\stackrel{\dots}{V} \\
h^{4}
\end{array}\right]=\underbrace{\left[\begin{array}{c}
v_1 \\
v_2
\end{array}\right]}_\text{Nominal linear part}+\underbrace{
\left[\begin{array}{c}
\Delta\stackrel{\dots}{V}(x,u,p) \\
\Delta {h}^4(x,u,p)\end{array}\right]}_\text{Uncertain nonlinear part}.
\end{equation}
\end{small}
Also, by using the fact that there are no uncertainty terms in $\dot{V}$, and $\dot{h}$, we can write linearized input-output map for the AHFV model using (\ref{eqyi}) as follows:
\begin{small}
\begin{equation}
\label{equyfblin}
\left[\begin{array}{c}
\dot{V}\\
\ddot{V}\\
\stackrel{\dots}{V} \\
\dot{h}\\
\ddot{h}\\
\stackrel{\dots}{h} \\
h^{4}
\end{array}\right]=\left[\begin{array}{c}
0\\
0\\
v_1 \\
0\\
0\\
0\\
v_2
\end{array}\right]+
\left[\begin{array}{c}
0\\
\Delta\ddot{V}\\
\Delta\stackrel{\dots}{V}(x,u,p) \\
0\\
\Delta\ddot{h}\\
\Delta\stackrel{\dots}{h}(x,u,p) \\
\Delta {h}^4(x,u,p)\end{array}\right].
\end{equation}
\end{small}
Let us define a diffeomorphism for each system as in (\ref{eqdiffm}) which maps the new vectors $\xi$ and $\eta$ respectively to the original vector $x$ as follows:
\begin{small}
\begin{equation}
\xi=T_1(x,p_0,V_c)
,\quad
\eta=T_2(x,p_0,h_c),
\end{equation}
where
\begin{align*}
&T_1(x,p,V_c)=\left[
\begin{array}{cccc}
\int_0^{t}{(V(\tau)-V_c)}d\tau & V-V_c & \dot{V} & \ddot{V}\end{array}\right]^T,\\
&T_2(x,p,h_c)=\left[
\begin{array}{ccccc}
\int_0^{t}{ (h(\tau)-h_c)}d\tau & h-h_c & \dot{h} & \ddot{h} & \stackrel{\dots}{h}\end{array}\right]^T,
\end{align*}
\end{small}
and $V_c$ and $h_c$ are the desired command values for the velocity and altitude respectively.
We write each diffeomorphism as follows:
\begin{small}
\begin{equation}
\label{eqtransformx}
\chi = T(x,p_0,V_c,h_c),
\end{equation}
where $\chi= \left[\begin{array}{ccccccccc}
\xi_1 & \xi_2 & \xi_3 &\xi_4 & \eta_1& \eta_2& \eta_3& \eta_4& \eta_5 \end{array}\right]^T$, and
$T(x,p,V_c,h_c)=\left[\begin{array}{cc}
T_1(x,p,V_c) \\
T_2(x,p,h_c) \end{array}\right]$.
\end{small}
Now, we can transform the nominal part of (\ref{equyfblin}) into the new states using the transformation (\ref{eqtransformx}) and linearize the uncertainty parts of (\ref{equfblin}) using the generalized mean value theorem  as follows:
\begin{small}
\begin{align}
\label{eqlfinalm}
\left[\begin{array}{c}
\dot{\xi}_1 \\
\dot{\xi}_2 \\
\dot{\xi}_3 \\
\dot{\xi}_4 \\
\dot{\eta_1}\\
\dot{\eta_2}\\
\dot{\eta_3}\\
\dot{\eta_4}\\
\dot{\eta}_5
\end{array}\right]=\left[\begin{array}{c}
0\\
0\\
0\\
v_1 \\
0\\
0\\
0\\
v_2
\end{array}\right]
+
\left[\begin{array}{c}
0\\
0\\
\Delta w_{1}(\tilde{\chi},\tilde{v},p) \\
\Delta w_{2}(\tilde{\chi},\tilde{v},p) \\
0\\
0\\
\Delta w_{3}(\tilde{\chi},\tilde{v},p) \\
\Delta w_{4}(\tilde{\chi},\tilde{v},p) \\
\Delta w_{5}(\tilde{\chi},\tilde{v},p)
\end{array}\right]\chi+
\left[\begin{array}{c}
0\\
0\\
0\\
\Delta \tilde{w}_{1}(\tilde{\chi},\tilde{v},p) \\
0\\
0\\
0\\
0\\
\Delta \tilde{w}_{2}(\tilde{\chi},\tilde{v},p)
\end{array}\right]v.
\end{align}
\end{small}
In this section, we write the equation (\ref{eqlfinalm}) in a structured form as presented in Subsection \ref{sec:structure_model}. Using (\ref{eqrhos}), (\ref{eqgforms}), and (\ref{eqGs}) we can write (\ref{eqlfinalm}) as given below:
\begin{small}
\begin{equation}
\label{eqgforms_ex}
\begin{split}
\dot {\chi}(t) &=(A+\tilde{C}_{1} \Delta_{1} \tilde{K}_{1}+\tilde{C}_{2} \Delta_{2} \tilde{K}_{2}+\cdots +\tilde{C}_{9} \Delta_{9} \tilde{K}_{9})\chi(t)\\
&+(B+\tilde{C}_{1} \Delta_{1} \tilde{G}_{1}+\tilde{C}_{2} \Delta_{2} \tilde{G}_{2}+\cdots +\tilde{C}_{9} \Delta_{9} \tilde{G}_{9}) v(t)
\end{split}
\end{equation}
\end{small}
where
\begin{footnotesize}
\[
\tilde{C}_3=\left[
\begin{array}{ccccccccc}
0 & 0 & 1 & 0 & 0 & 0 & 0 & 0 & 0
\end{array}\right],\]
\[\tilde{C}_4=\left[
\begin{array}{ccccccccc}
0 & 0 & 0 & 1 & 0 & 0 & 0 & 0 & 0
\end{array}\right],
\]
\[
\tilde{C}_7=\left[
\begin{array}{ccccccccc}
0 & 0 & 0 & 0 & 0 & 0 & 1 & 0 & 0
\end{array}\right],\]
\[\tilde{C}_8=\left[
\begin{array}{ccccccccc}
0 & 0 & 0 & 0 & 0 & 0 & 0 & 1 & 0
\end{array}\right],
\]
\[
\tilde{C}_9=\left[
\begin{array}{ccccccccc}
0 & 0 & 0 & 0 & 0 & 0 & 0 & 0 & 1
\end{array}\right],\]\[
\tilde{C}_1=\tilde{C}_2=\tilde{C}_5=\tilde{C}_6=\left[
\begin{array}{ccccccccc}
0 & 0 & 0 & 0 & 0 & 0 & 0 & 0 & 0
\end{array}\right];
\]
\[
\tilde{K}_3=\left[
\begin{array}{ccccccccc}
\hat{\rho}_3 & \hat{\rho}_3 & \hat{\rho}_3 & \hat{\rho}_3 & \hat{\rho}_3 & \hat{\rho}_3 & \hat{\rho}_3 & \hat{\rho}_3 & \hat{\rho}_3
\end{array}\right],\]
\[\tilde{K}_4=\left[
\begin{array}{ccccccccc}
\hat{\rho}_4 & \hat{\rho}_4 & \hat{\rho}_4 & \hat{\rho}_4 & \hat{\rho}_4 & \hat{\rho}_4 & \hat{\rho}_4 & \hat{\rho}_4 & \hat{\rho}_4
\end{array}\right],
\]
\[
\tilde{K}_7=\left[
\begin{array}{ccccccccc}
\hat{\rho}_7 & \hat{\rho}_7 & \hat{\rho}_7 & \hat{\rho}_7 & \hat{\rho}_7 & \hat{\rho}_7 & \hat{\rho}_7 & \hat{\rho}_7 & \hat{\rho}_7
\end{array}\right],\]
\[\tilde{K}_8=\left[
\begin{array}{ccccccccc}
\hat{\rho}_8 & \hat{\rho}_8 & \hat{\rho}_8 & \hat{\rho}_8 & \hat{\rho}_8 & \hat{\rho}_8 & \hat{\rho}_8 & \hat{\rho}_8 & \hat{\rho}_8
\end{array}\right],
\]
\[
\tilde{K}_9=\left[
\begin{array}{ccccccccc}
\hat{\rho}_9 & \hat{\rho}_9 & \hat{\rho}_9 & \hat{\rho}_9 & \hat{\rho}_9 & \hat{\rho}_9 & \hat{\rho}_9 & \hat{\rho}_9 & \hat{\rho}_9
\end{array}\right],
\]
\[
\tilde{K}_1=\tilde{K}_2=\tilde{K}_5=\tilde{K}_6=\left[
\begin{array}{ccccccccc}
0 & 0 & 0 & 0 & 0 & 0 & 0 & 0 & 0
\end{array}\right];
\]
\[
\tilde{G}_4=\left[
\begin{array}{cc}
\hat{\rho}_4 & \hat{\rho}_4
\end{array}\right],\]
\[\tilde{G}_9=\left[
\begin{array}{cc}
\hat{\rho}_9 & \hat{\rho}_9
\end{array}\right],
\]
\[
\tilde{G}_1=\tilde{G}_2=\tilde{G}_3=\tilde{G}_5=\tilde{G}_6=\tilde{G}_7=\tilde{G}_8=\left[
\begin{array}{cc}
0 & 0
\end{array}\right].
\]
\end{footnotesize}
Using the above definition of the variables, we can write the system in the general MIMO form given in (\ref{eqggforms}), where $\chi(t)\in \mathbb{R}^9$ is the state, $\zeta_k (t)=\Delta_{k}[\tilde{K}_k\chi+\tilde{G}_1 v] \in \mathbb{R}$ is the uncertainty input, $z_k(t)\in \mathbb{R}$ is the uncertainty output, and $v(t)\in\mathbb{R}^2$ is the new control input vector.
\subsection{Minimax LQR Control Design}
The linearized model (\ref{eqggforms}) corresponding to the AHFV uncertain nonlinear model (\ref{eqn2}) allows for the design of a minimax LQR controller for the velocity and altitude reference tracking problem. The method of designing a minimax LQR controller is given in \cite{IP}. Here, we follow the same method and proposed a minimax LQR controller for the linearized system (\ref{eqggforms}). We assume the uncertainty in the system (\ref{eqggforms}) satisfies following IQC and the original state vector $x$ is available for measurement. 
\begin{small}
\begin{equation}
\label{eqIQC}
\int_0^{\infty}(\parallel z_j(t)\parallel^2-\parallel \xi_j(t)\parallel^2)dt\geq-\chi^T(0)D_j\chi(0),
\end{equation}
\end{small}
where $D_j>0$ for each $j=1,\cdots,m$ is a given positive definite matrix. The cost function selected is as given below:
\begin{equation}
\label{eqfcost}
F=\int_0^{\infty}{[\chi(t)^T Q \chi(t)+v(t)^T R v(t)]dt},
\end{equation}
where $Q=Q^T>0$ and $R=R^T>0$ are the state and control weighting matrices respectively. A minimax optimal controller can be designed by solving a game type Riccati equation
\begin{small}
\begin{align}
\label{eqARE}
&(A-BE^{-1}G^T K)^T X_\tau +X_\tau(A-BE^{-1}G^T K)\nonumber\\&+X_\tau(CC^T-BE^{-1}B^T)X_\tau
+K^T(I-GE^{-1}G^T)K=0,
\end{align}
where
\[
K=\left[\begin{array}{c}
Q^{1/2} \\
0 \\
\sqrt{\tau_1}\tilde{K}_1 \\
\vdots\\
\sqrt{\tau_9}\tilde{K}_9
\end{array}\right]
,\quad
G=\left[\begin{array}{c}
0 \\
R^{1/2} \\
\sqrt{\tau_1}\tilde{G}_1 \\
\vdots\\
\sqrt{\tau_9}\tilde{G}_9
\end{array}\right]
,\quad
E=GG^T,
\]
\[
C=\left[\begin{array}{ccc}
\frac{1}{\sqrt{\tau_1}}C_1 & \dots & \frac{1}{\sqrt{\tau_9}}C_9
\end{array}\right]
\]
\end{small}
The weighting matrices $Q$ and $R$, and parameters $\tau_k$, for $k=1,2,\cdots,9$ are selected such that they give the minimum bound
\begin{small}
\begin{equation}
\label{eqfbond}
\text{min}[\chi^T(0)X_\tau \chi(0)+\sum_{j=1}^{9}\tau_j \chi^T(0)D_j\chi(0)],
\end{equation}
\end{small}
on the cost function (\ref{eqfcost}). The minimax LQR control law can be obtained by solving the ARE (\ref{eqARE}) for given values of the parameters as given below:
\begin{small}
\begin{equation}
\label{eqcontrllaw_mv}
v(t)=-G_\tau\chi(t),
\end{equation}
where
\[
G_\tau=E^{-1}[B^T X_\tau+G^T K]
\]
\end{small}
is the controller gain matrix. The parameters $Q$ and $R$ are selected intuitively so that required performance can be obtained and $\tau_k$ for $k=1,2,\cdots,9$ correspond to the minimum bound on (\ref{eqfcost}). These parameters are given as follows:
\begin{small}
\begin{equation*}
\label{eqSW_mv}
\mathbf{Q}=\textbf{diag}\left[
\begin{array}{c}
1000, 500, 500, 100, 0.001, 100, 100, 500, 500
\end{array}\right],
\end{equation*}
\begin{equation*}
\label{eqCW_mv}
\mathbf{R}=\left[
\begin{array}{cc}
3.0 & 0\\
0 & 3.0
\end{array}\right],\quad \tau_3=547.9,
\tau_4=8.0,~~ \tau_7=4935.7,
\end{equation*}
\[
\tau_8=4935.3,~~\tau_9=3768.0.
\]
\end{small}
\subsection{Simulation Results} \label{sec:resultslqr_mv}
\begin{figure}[htbp]
\hfill
\begin{center}
\epsfig{file=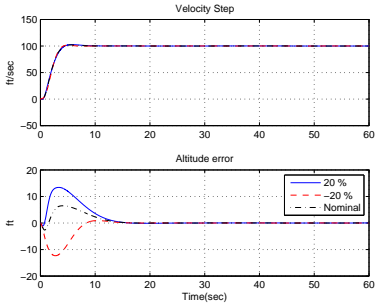, scale=0.38}
\caption{System response to a step change in velocity reference.}
\label{fig:vstep_mv1}
\end{center}
\end{figure}

\begin{figure}[htbp]
\hfill
\begin{center}
\epsfig{file=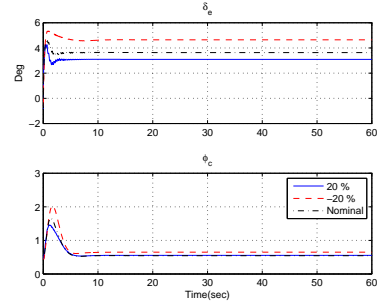, scale=0.38}
\caption{Control input responses corresponding to a step change in velocity reference.}
\label{fig:vstep_mv}
\end{center}
\end{figure}

\begin{figure}[htbp]
\hfill
\begin{center}
\epsfig{file=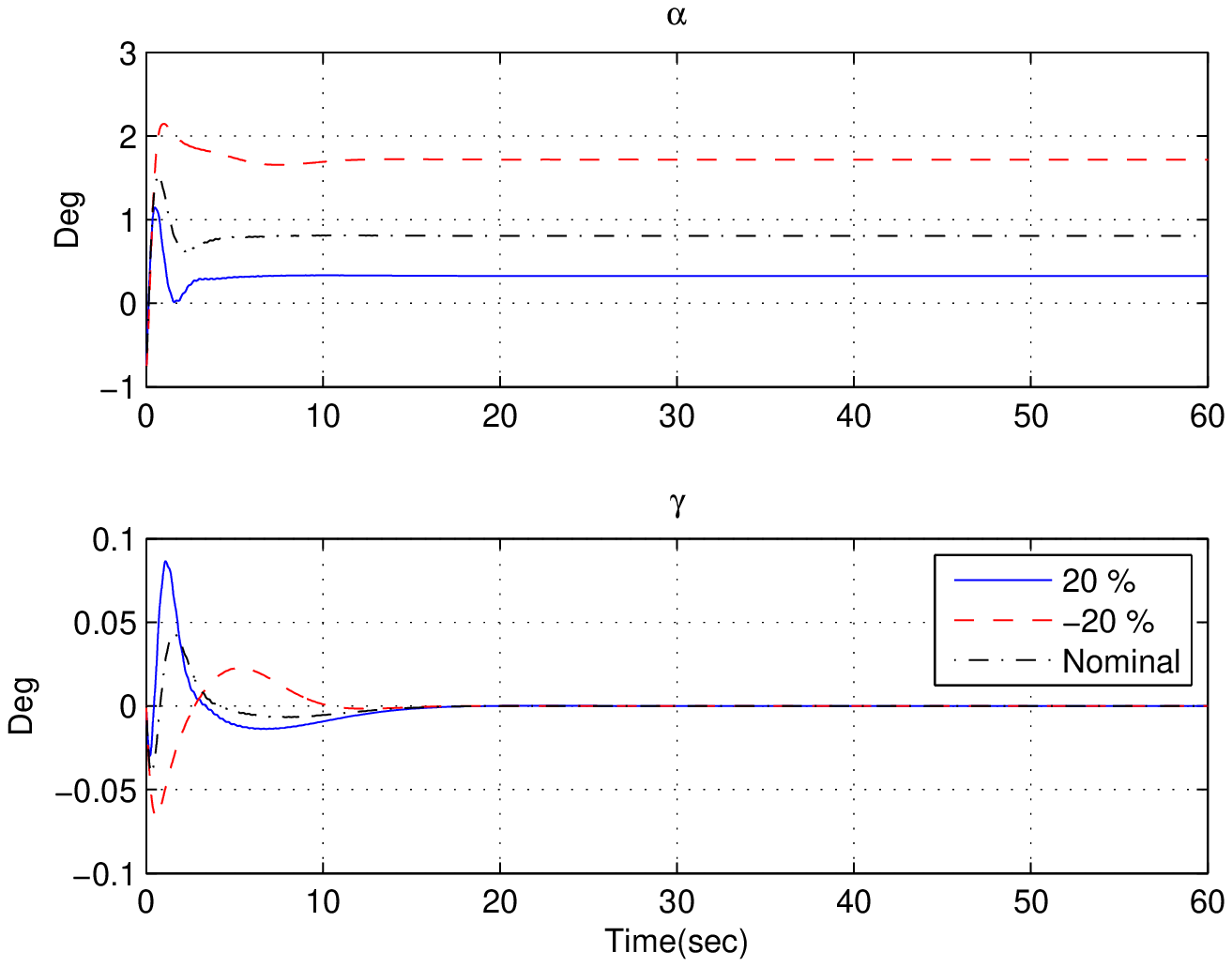, scale=0.38}
\caption{Responses of $\alpha$ and $\gamma$ for a step change in velocity reference.}
\label{fig:vstep_states_mv1}
\end{center}
\end{figure}

\begin{figure}[htbp]
\hfill
\begin{center}
\epsfig{file=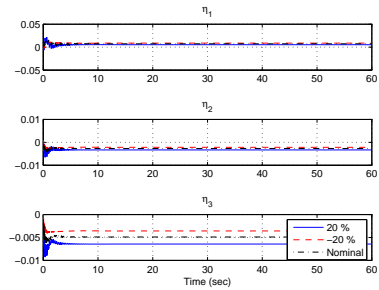, scale=0.38}
\caption{Flexible states during velocity reference tracking.}
\label{fig:vstep_flex_mv}
\end{center}
\end{figure}

\begin{figure}[htbp]
\hfill
\begin{center}
\epsfig{file=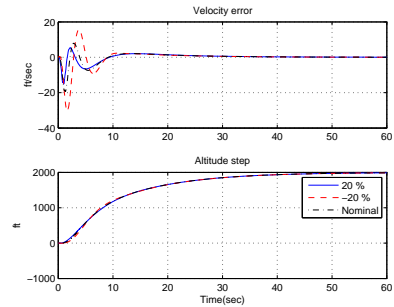, scale=0.38}
\caption{System response to a step change in altitude reference.}
\label{fig:hstep_mv2}
\end{center}
\end{figure}

\begin{figure}[htbp]
\hfill
\begin{center}
\epsfig{file=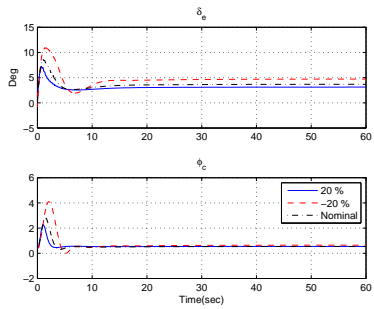, scale=0.38}
\caption{Control input responses corresponding to a step change in altitude reference.}
\label{fig:hstep_mv}
\end{center}
\end{figure}

\begin{figure}[htbp]
\hfill
\begin{center}
\epsfig{file=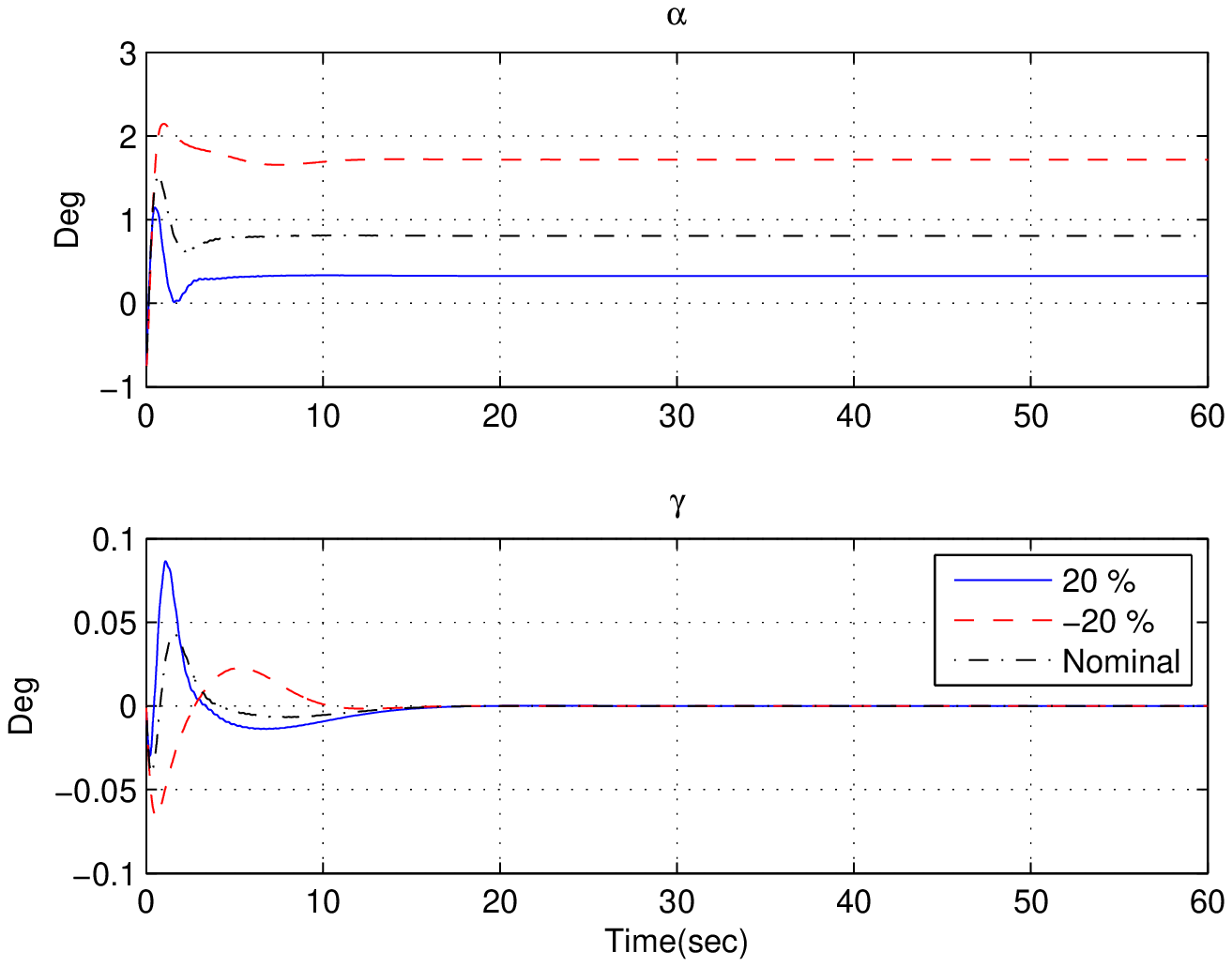, scale=0.38}
\caption{Responses of $\alpha$ and $\gamma$ for a step change in altitude reference.}
\label{fig:hstep_states_mv3}
\end{center}
\end{figure}

\begin{figure}[htbp]
\hfill
\begin{center}
\epsfig{file=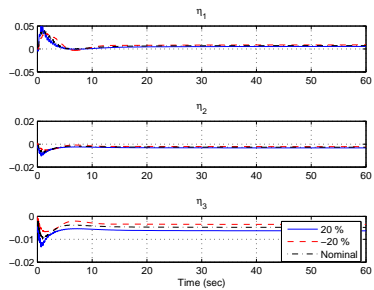, scale=0.38}
\caption{Flexible states during altitude reference tracking.}
\label{fig:hstep_flex_mv}
\end{center}
\end{figure}

The closed loop nonlinear AHFV system with the minimax LQR controller (\ref{eqcontrllaw_mv}) is simulated using different sizes and combinations of uncertainty. For the sake of brevity, here we evaluate the performance of the proposed controller by using step input commands for the following three cases:
\begin{itemize}
\item[1.] Uncertain parameters equal to their nominal values, with no uncertainty.
\item[2.] Uncertain parameters $20\%$ lower than their nominal values.
\item[3.] Uncertain parameters $20\%$ larger than their nominal values.
\end{itemize}
The responses of the closed loop system given in Fig. \ref{fig:vstep_mv1} -- Fig. \ref{fig:hstep_flex_mv} show that the minimax LQR controller along with the feedback linearization law gives satisfactory performance.
\section{Conclusion}\label{sec:concl}
In this paper, a robust nonlinear tracking control scheme for a class of uncertain nonlinear systems has been proposed. The proposed method uses a robust feedback linearization approach and the generalized mean value theorem to obtain an uncertain linear model for the corresponding uncertain nonlinear system. The scheme allows for a structured uncertainty representation.  In order to demonstrate the applicability of the proposed method to a real world problem, the method is applied to a tracking control problem for an air-breathing hypersonic flight vehicle. Simulation results for step changes in the velocity and altitude reference commands show that the proposed scheme works very well in this example and the tracking of velocity and altitude is achieved effectively even in the presence of uncertainties.

\end{document}